# Platform-as-a-Service (PaaS): The Next Hype of Cloud Computing

## Robail Yasrab

*School of Computer Science and Information Technology,*
*University of Science and Technology China (USTC),*
*Hefei China.*

## Abstract

Cloud Computing is expected to become the driving force of information technology to revolutionize the future. Presently number of companies is trying to adopt this new technology either as service providers, enablers or vendors. In this way the cloud market is estimated be likely to emerge at a remarkable rate. Under the whole cloud umbrella, PaaS seems to have a relatively small market share. However, it is expected to offer much more as it is compared with its counterparts SaaS and IaaS. This paper is aimed to assess and analyze the future of PaaS technology. Year 2015 named as "the year of PaaS". It means that PaaS technology has established strong roots and ready to hit the market with better technology services. This research will discuss future PaaS market trends, growth and business competitors. In the current dynamic era, several companies in the market are offering PaaS services. This research will also outline some of the top service providers (proprietary & open source) to discuss their current technology status and present a futuristic look into their services and business strategies. Analysis of the present and future PaaS technology infrastructure will also be a major discussion in this paper.

## 1 Introduction

In the recent past, most of the IT oriented people were thinking that the cloud computing paradigm is only a Software-as-a-Service (SaaS) and Infrastructure-as-a-Service (IaaS). However, after the release of Heroku, Windows Azure, AWS (Amazon web Services), and Google App-Engine people have become more and more well known to Platform-as-a-Service (PaaS) [1]. Currently, a number of companies across the world have moved their technology development practices and processes to the cloud through PaaS to transform the traditional development approaches and to get the full benefit from state of the art flexible services. However, these developments have not been beneficial to PaaS as compared to the popularity gained by IaaS and SaaS cloud services.

Gap between IaaS and SaaS has been filled by PaaS and it allows application owners to still own and implement their own applications without depending on basic IT infrastructure [2]. For instance, a software developer is not required to worry about installing, monitoring, configuring, load balancing, patching or some other issues related to runtime environment and middleware when on PaaS platform [3]. According to specifications of the given cloud platform, a developer is just required to deploy the application online.

PaaS infrastructure offers a number of IT and business based benefits, e.g. it offers more business agility to an enterprise than a traditional software development approach. The development of an application "Born on PaaS" offers better TTV (Time to Value) factor because such applications can be developed so quickly and can be deployed so faster [4]. The application "Born on PaaS" minimize the time for moving parts of infrastructure. This means it consumes less time and cost resources. It also offers a highly standardized scalable delivery and self-managing platform. This is a massive benefit for the enterprise application deployment and development.

The emergence of PaaS cloud market has given direction to a lot of major proprietary players like, Amazon Web Services (AWS), Google, Salesforce.com, Heroku and Microsoft's Azure to present theirselves on cloud market. Every IT company is struggling for the top position in the PaaS cloud market. Choosing a vendor for PaaS services is not easy because companies like Cloud Foundry and OpenShift of Red Hat are also trying to establish solid ground in the open source PaaS market. The proprietary PaaS vendors are also performing excellent in the cloud market which again has created a great platform rivalry.

Currently, many PaaS vendors are in market and offering sophisticated cloud services, though still PaaS lagging is behind as compared to SaaS and IaaS. Most of the organizations worldwide are hesitant to adopt private PaaS solutions, due to numerous security and privacy concerns. Currently, many of security and privacy issues are neutralized. However, is it enough to enhance PaaS' worldwide extensive adoption? There are a lot of issues and factors who are presently hindering the widespread adoption of PaaS. This research will try to find the answer of the above question and will address the certain aspects of PaaS, e.g. emerging trends, market forecasts, market competition and future services. The first section of this research paper is outlined with the current status of PaaS market. Second section is based on PaaS marker futuristic analysis. Third section represents the analysis and discussion of key market players of PaaS and the last section is about conclusion and work sighted.

## 2 PAAS' Present

PaaS technology was first introduced by Salesforce.com, Heroku and Amazon Web Services (AWS) back in year 2007. After this Google launched App Engine in 2008 which was a free trial version. Overnight, these initiatives transformed the online cloud computing space into a complete virtual industry. At that time PaaS was not introduced as an extension of SaaS. It was offered as a complete solution for enterprise customized Research and Development (R&D) middleware platform. Now PaaS technology has evolved and not only offers application software but also covers both application and database services. PaaS cloud technology has changed the development, testing and deployment processes of service-oriented architecture (SOA). It has also provided ease in the application development process of whole SOA architecture [5].

Presently, Platform-as-a-Service platforms are implemented in different styles. PaaS Service providers acquire different approaches to offer better services. One of the most popular categories is "SaaS Environment Anchored PaaS". Such PaaS cloud vendors have SaaS as their core business services. To grow their business capabilities and brand, the PaaS vendors develop an ecosystem that permits independent software developers to offer applications on top of the cloud platform.

The next category is "Operating Environment Anchored" PaaS, where PaaS is tied to an operation environment that makes it easier to operate and perform specific actions within that environment. Such PaaS vendors offer the entire foundations, e.g. the OS, networking, storage and so on. Moreover, it offers the tools for application development.

The third category is "Open PaaS" that is aimed to encourage an open process and environment which is not tied to a particular cloud computing implementation. Open PaaS providers permit developers to bring their own platform to the cloud which leads to flexibility. However, it can add more cost and complexity [6].

More and more businesses worldwide have recognized the significance of PaaS technology solutions. Though, businesses are confused between adoptions of PaaS technology. PaaS technology is available in different flavors, like private PaaS (hosted in-house) or public PaaS cloud. Private PaaS demand extra set-up cost whereas public PaaS cloud system has security, privacy and other issues [6]. Transition between private and public PaaS is not so easy. So the solution is Hybrid PaaS cloud. It is a blend of private and public PaaS cloud. Hybrid PaaS offers more businesses and performance flexibility.

Currently, the cloud market is filled with the number of PaaS service providers. Each company is trying to capture the maximum market share with maximum cloud services. So, the number of companies has started offering PaaS services beside the established IaaS or SaaS market. Most of the companies have already converged on this space. Like, Amazon's Elastic Beanstalk or the Relational Database Service moves up from the IaaS layer, while some companies like Salesforce's Force.com which moves down from the SaaS layer. But on the other hand, some companies took new initiative, like Google with App-Engine and Microsoft with Windows Azure. However, there are also some pure players who are offering pure PaaS services like EngineYard [7].

Furthermore, in near future we are expecting more and more new companies investing and initiating PaaS services due to its potential high market acceptance rate. Another reason is the role of open standards as well as the open source of PaaS services. In contrary to IaaS where the Amazon has developed a bit of a de-facto standard, PaaS is probably turned out to be more and more de-jure (open standards). IBM recently declared PaaS capability named "BlueMix" that is a $1 billion project. While based on open source standards RedHat's "OpenShift" PaaS platform is also turning out to be a great active future contributor to PaaS [7].

Bal et al. [8] stated that PaaS has transformed the ways of dealing new technology challenges like web based computing, global software development, open-source development and outsourcing of the major business process. This smooth application development process offered a great deal of ease for the people who involved with development of WOA (web-oriented architecture) and SOA. Because now they are able to decide where to host services and processes either outside or inside of the firewall system.

PaaS offered extremely low cost, efficient and large-scale application development on the Internet. Though PaaS not only promote the worth of speed and cost of development, but also enhance the web based resource utilization[8]. It is believed that soon PaaS will emerge as a new era of rapid development cloud platform.

PaaS cloud services offer a great deal of freedom and productivity to developers. Because they don't need to worry about defining scalable requirements also they do not need to use XML to outline the specification of the system deployment. All these facilities are offered by the PaaS providers with responsibility to manage all these services. According to Grover & Kheterpal [10], PaaS offers services for developers like application design, development, testing and deployment. In addition to secondary services needed for development; like web-service integration, team-collaboration, security, database integration, storage, scalability, state management, persistence, application instrumentation, versioning, and developer community facilitation. These all services offered by present PaaS technology platforms. Other non-engineering services are also offered for development management; like project monitoring, project workflow management, tools for service management, discovery and reservation handling.

Besides all these merits, the major issue with PaaS cloud technology is that developers are being "locked-in" to a certain platform. For instance, a company has taken services of Google App Engine for their applications and wants to move to Amazon' PaaS cloud; in this case the company has to completely re-write their applications[9]. These days most of the new PaaS vendors offering "lock-in free" services. In PaaS 1.0 cloud "Lock-in" was introduced as a feature. However, due to the new era of higher global collaboration, out-sourcing, open source development and interoperability; there is no excuse for offering vendor lock-in feature. So, PaaS 2.0 cloud is open by design, offering more freedom and high sharing [5, 10].

Cloud security always been one of the hottest debates since its emergence. Nowadays, cloud vendors normally offer improved protection methods because it's in the interest of vendors to offer high level of security for their cloud clients[11, 12]. PaaS normally offers a relatively refined suite of access controls. Though clients are not having any risk but vendors still own the risk.

## 3 Future Market Analyses

IaaS and SaaS market investment is on the rise but PaaS technology sector tends to lag behind. As different companies are showing their concern about cloud privacy, security and management concerns; the PaaS vendors are hastening to defeat those concerns and fears [15]. This study leads to following questions. 1) Can private PaaS cloud (more secure PaaS) option help to boost adoption in the market? 2) Or other options, like IaaS cloud vendors capitalizing the cloud market and delivering PaaS platform services from the same platform? This section is focused on answering these questions and analysis of the future market size of PaaS among other cloud services (IaaS and SaaS platforms) and regional markets.

InfoWorld is worldwide well accepted technology news agency and part of the International Data Group. This news agency is one of the best choices for business management, technology decision makers and leaders seeking for expert analysis of enterprise technology [16]. According to a report on the PaaS technology future, it is predicted that PaaS cloud industry will rise rapidly due to great interest of global software technology developers in faster, easier and flexible application development with nominal IT budget [15].

International Data Corporation (IDC) is another major international leading organization for technology advisory services, technology market intelligence and IT analyses has predicted the future of PaaS on the basis of PaaS 2011–2013 revenue. IDC forecasted 2014–2018 international competitive public cloud PaaS market. It is stated that a number of prospects for monetizing public PaaS technology is continue to rise through growing IT solutions, approximately from all global businesses. The main reason is the growth of data from mobile and social communication along-side events being gathered from the "Internet-of-Things". All of these initiatives are now tightly connected to the cloud and turning to be a massive force that demanding public PaaS and improving public PaaS potentials [17].

According to Kanaracus [18] worldwide market of PaaS technology is going to set a big leap. It is expected that, the US \$3.8 billion market in 2012, jump to \$14 billion market till 2017. This big leap is expected because worldwide businesses want to reduce infrastructure costs as well as enhance the speed of application development and deployment [18] & IDC. The total PaaS annual growth rate according to present situations will be roughly 30%, according to InfoWorld.

As IDC predicted 32% growth in the public PaaS cloud market since 2013 and claimed a thrilling future for global public PaaS. It is also stated that the Public PaaS market will grow from 13% to 16% from overall cloud market revenue till 2018 [17]. It demonstrates that it will continue to grow to become an influential tool to facilitate clients and to meet the growing market and business needs. According to another report by MarketResearch.com "Global PaaS Market 2010-2014" the worldwide PaaS market is predicted to be raised up-to 26% annually through 2014 [19,20].

At Gartner "Application Architecture, Development & Integration Summit, 2011" outlined some of the noticeable figures regarding the PaaS technology future. According to report, it was stated that till 2015, enterprise adoption of PaaS will rise from 3% to 43%. [21-23].

From overall geographic PaaS market revenue point-of-view, highest revenue generated from the USA is 65.2% in 2012. It is also predicted that this figure is not going to have big change up-to 2017. The major reason behind the stable revenue is that the several initial PaaS vendors have their roots in USA and offering such "revenue biasness". However, PaaS market revenue expansion in the Asia-Pacific including Japan is "continue to expand" by 14.1% market share in 2012 and it is expected to expand to 19.0% up-to 2017. While the Middle East, Europe and Africa collectively share 20.7% PaaS annual revenue in 2012 but with a slight drop in revenue up-to 18.7% by 2017 as shown in Fig.1 [18].

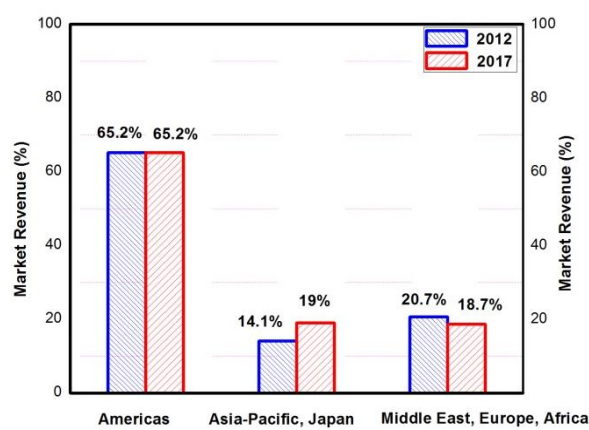

Fig. 1 Geographic PaaS market (2012-2017)

According to Visiongain forecast for the PaaS submarket, revenue is valued at $1.9 Billion in the year 2013 and it is expected to increase $3.7 Billion till year 2018 [24]. Fig.2 demonstrates the Visiongain's forecast for PaaS till 2018.

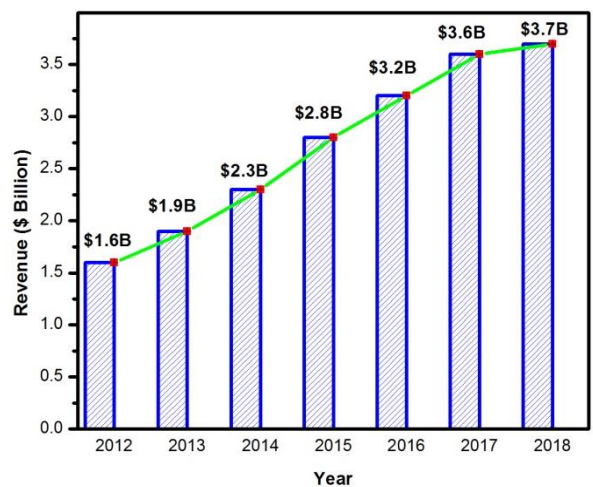

**Fig. 2** PaaS Market forecast 2013-2018

Fig.3 shows the annual growth rate of PaaS technology from year 2012 to 2018. This annual growth rate is directed related to above stated Visiongain's forecast annual increase in revenue. It can be clearly seen that year 2014 and 2015 are the peak development years for PaaS technology.

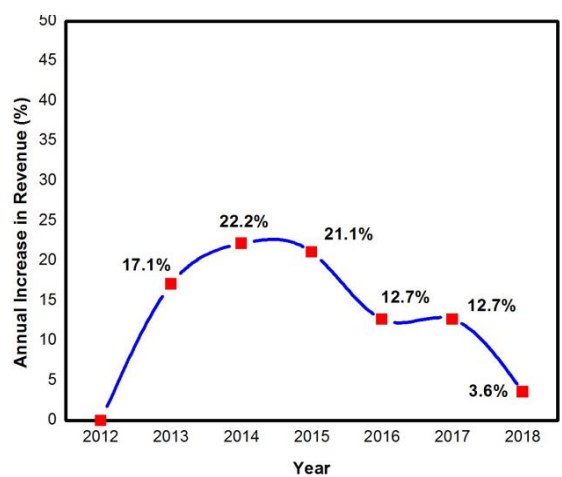

**Fig. 3** PaaS Annual Growth Rate (%) 2012-2018

"Compound Annual Growth Rate (CAGR) is defined as the year-over-year growth rate of an investment over the given time period". It is an investment and business related term for the geometric progression ratio that offers a constant rate of return over the specific time period [25]. According to Gartner, Public Cloud Services Forecast 3Q13, September 2013, enterprises will spend $921 billion on public cloud services till 2017 with overall 17% CAGR (2012-2017) as shown in the Fig.4. This forecast of Gartner offers public cloud future's preview. Also according to Visiongain forecast, the PaaS cloud is going to attain a 14.3% CAGR for the period 2013-2018

[24]. If we assess these figures closely, it is obvious that all these forecasts are approximately offering same numbers.

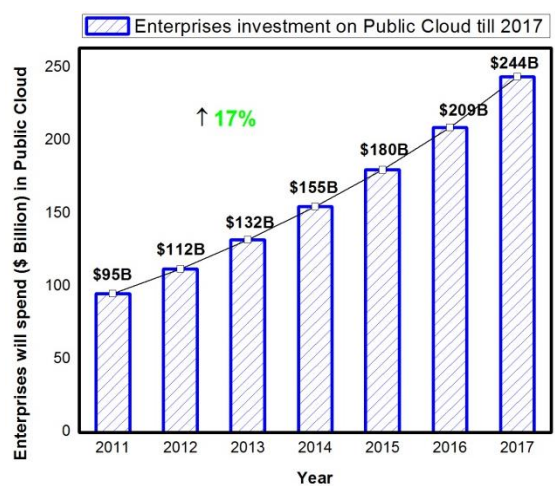

**Fig. 4** Enterprises investment on Public Cloud till 2017

"North Bridge Venture Partners" in collaboration with "GigaOM Research" and 57 collaborating organizations performed a future analysis survey of cloud services that offers a glimpse into the cloud computing technology adoption trends, inhibitors, and driving forces for long-term growth. The results of this survey outlined cloud segments and general growth analysis which are forecasted from 451-Research Market Monitor Report [26]. The Fig.5 given below outlines the CAGRs and worldwide addressable market capacity by IaaS, PaaS and SaaS. It shows the comparison of 2012 results and 2016 market forecasts. This graph clearly depicts that PaaS is ultimately having the highest growth rate and CAGR in the worldwide market.

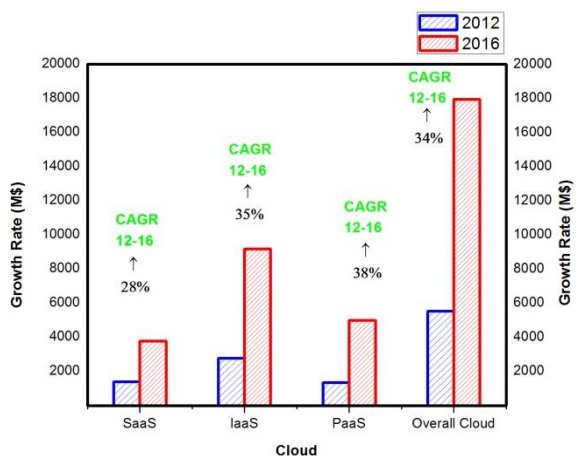

**Fig. 5** SaaS, IaaS & PaaS growth rate and CAGR comparison (2012-2016)

According to 451-Research analysis report on the PaaS, future market will attain 38% CAGR through 2016 by producing 24% of overall cloud computing services revenue. It is also assessed that 71% of PaaS revenue will be produced through vendors over $75M in sales [24]. Another 451-Research report on "The future of cloud services" forecasted that PaaS is a fastest growing industry of the future. Hosted Infrastructure Services (current growth rate 69%) and SaaS (current growth rate 71%) are two popular cloud services. Having the future forecasted growth rate of 14% for both IaaS and SaaS by 2016. With 37% current growth rate, PaaS is projected to have highest annual growth rate that is 26 upto 2016 as shown in Fig.6. This demonstrators that PaaS is turning out to be one of top revenue generator in cloud family [27].

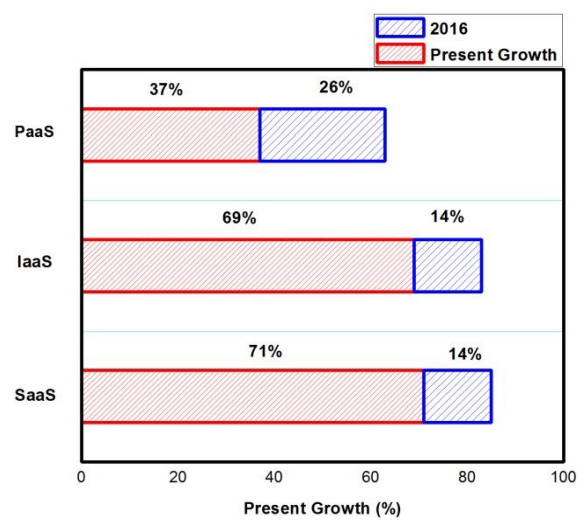

**Fig.6** PaaS, IaaS & SaaS future growth analysis (Current-2016)

# 4. PaaS: Developers, Operations and Service Providers Future

Eventually, the objective of any PaaS technology is to ease application developers to develop and implement applications. Also the other objective of PaaS is to make sure that the primary services interact mutually and with core hardware infrastructure in a right way. Such aspects should be assessed before adopting PaaS cloud whether the PaaS cloud service provider is a public PaaS vendor (like AWS or Azure) or an independent entity (like dotCloud, Heroku or AppFog) [28].

According to Gartner, the year 2015 is declared as the year of PaaS because this technology is influencing developers, operations and providers in an effective way. So, it is completely true that PaaS represents the future of development and IT. Let's have a look how PaaS will transform and facilitate the functioning and performance of the PaaS developers, operations and providers in the future.

From developer's viewpoint, PaaS offers freedom to code low level plumbing as part of the overall application development process. According to Keller & Rexford [29], developers will make use of PaaS packages for code plumbing. Instead of wasting time in writing codes to monitor resource load and to start and stop virtual machines to manage the application loads, developers of PaaS can focus on valued part of the application and engineering business functionality. Currently developers are able to control pre-configured application settings which facilitates the quick development of systems through complex deployment of topologies. All these services offer a win-win situation for developers for improving their productivity.

From operation's viewpoint, PaaS bridges the challenge of shadow IT. Now application developers directly obtaining cloud resources. State-of-the-art PaaS technology solutions contain the development and operational tools like logging, monitoring and version controls thus, it streamline the operational environments. These all services and support offers a great deal of capability to enhance the developer performance and reduction in application development and deployment time.

The PaaS market is virtually divided into a number of technology platforms and services. Each service and platform offers different levels of flexibility, productivity and quality. These PaaS services also differ in capability and performance. PaaS is now evolving to new generation of services. One of most popular category is the "Application platform as a Service (aPaaS)". aPaaS offers a variety of capabilities like development tools, container service programming models and much more. In the present technological advanced era, these elements of aPaaS have gradually begun appearing in the marketplace. PaaS vendors like Google App Engine, Force.com, Cloud Foundry, Heroku, Engine Yard and Microsoft Azure have initiated some early adoption aPaaS technology. aPaaS has established the foundation on which numerous well known SaaS applications have already been built and much popular in these days.

iPaaS is the next most popular type of PaaS technology that is evolving as the next generation integration platform. This is designed for integrating cloud applications with one another. iPaaS also allows the application integration with on-premises and legacy applications. It is stated explicitly, an iPaaS is cloud integration PaaS which allows connectivity to SaaS as well as cloud services. Also it offers a safe method of accessibility for on-premises applications [30].

In past, PaaS services were offered in a dis-integrated fashion. However, now PaaS turn out to be a complete package of all customized services. Now technology providers are offering all these services in a single package. It is also expected that in near future, this technology paradigm will incorporate more and more specialized tools and technologies under the umbrella of PaaS. Fig.7 shows the evolution of PaaS services and integration of these services to a single package.

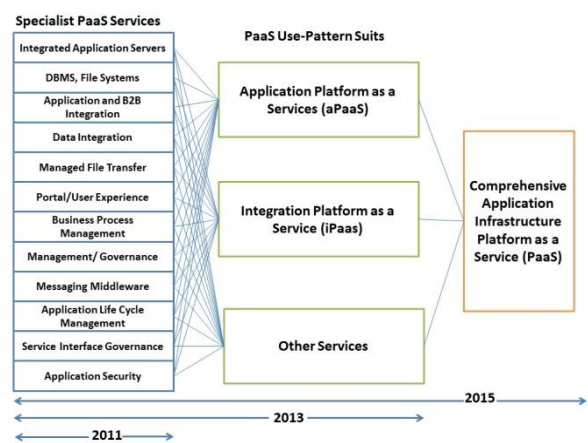

**Fig. 7** PaaS Roadmap

In addition to adoption of PaaS services, there is also need to know about most popular PaaS services and reason to adopt these services in future. In a report by Gartner, "Market Trends: Platform as a Service, Worldwide, 2012-2016, 2H12" published on Oct 5, 2012, it was stated that Application Platform-as-a-service (aPaaS) is a most popular PaaS service. According to the above mentioned Gartner forecast report, "Application Platform Services (aPaaS)" was the largest segment of the PaaS market producing 35% of total PaaS earnings in 2011 as shown in the Fig.8. Fig-8 shows the proportion of different PaaS services and future growth (201 to 2016) of these services in PaaS paradigm [31]. It also shows that in future aPaaS will be the preferred technology for companies.

willing to absorb losses. Prices also depend on the uniqueness and quality of services offered by any provider. Quality-of-services (QoS) is turning out to be a key aspect of cloud acceptance. Now cloud providers are struggling to offer more and more efficient services to grab maximum market share. Quality of service involves number of aspects like efficient load balancing, completion time improvement, response time optimization, reduction in bandwidth wastage, make-span improvement and accountability of the overall system [33].Quality aware PaaS providers are able to offer value based services related to application functionality. Those eventually linked to business productivity outcomes. This becomes a reason that businesses are willing to pay a premium. This also offers a very bright future for PaaS service provider

**Fig. 8** PaaS Market 2011 [Source: Gartner, Market Trends: Platform as a Service, Worldwide, 2012-2016, 2H12]

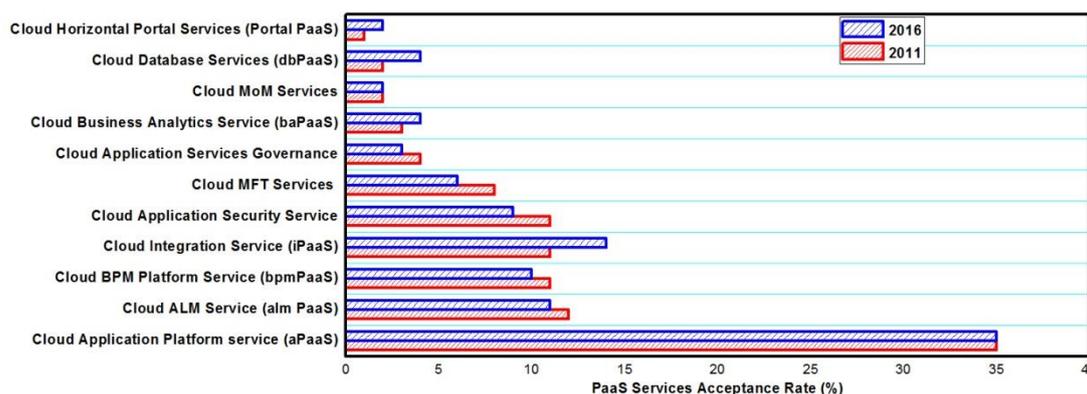

"Application Platform as a Service (aPaaS) is best defined by the Gartner. It says that aPaaS is a cloud service that offers development & deployment architecture for application services" [32]. It means that aPaaS is offering the Platform as a Service along with SaaS and IaaS. It offers end users with operating systems, hardware, network or storage capacity and other services to run existing applications or develop new ones. The most popular aPaaS service is Amazon AWS for PaaS. It is highly popular because it offers more flexibility for developers, improve productivity and enhance the quality of development. The Fig.8 shows the past and future trend of PaaS services. It also shows that aPaaS will be constantly occupying a major PaaS market share in the future. Here highly evolving trend of aPaaS directs to the major technology provider of PaaS in the coming future.

In the PaaS market, prices are set and managed by service providers who are most competent or where they can provide higher-margin services and differentiation to create a better competitive edge and earn more revenue [34].

According to Peng, et al. (2009) rather supporting developers, operations and providers, PaaS is now breaking away from its legacy and performing as a primary tool for cost and margin management for technology development firms [35]. In recent times, PaaS has offered the cloud to initiate technology evolution which is known as "virtual cloud operating system". Altogether, every time a technology vendor offers a fresh cloud service. It has effectively developed a virtual cloud OS for PaaS. Nowadays, the majority of PaaS services reflect a traditional OS. However, in near future, different PaaS technology providers' infrastructure will differ from each other and from current operating system to develop the real cloud-based operating system.

In connection to adoption PaaS technology by developers and development organization; a survey, "Develop Your Own Application (DYOA)" by "Business-Wire" carried out in June 2014. The basic objective of this survey was to assess the basic PaaS technology aspects and trends in current software industry. This research study outlined that PaaS technologies are well satisfying the rising demand of rapid application development and changing business needs. This survey took the views of 700 IT industry decision makers from all over the world (UK, US, Germany, France, Australia, Benelux, Singapore and Brazil). It is assessed that there is a huge demand of PaaS technologies from organizations those are searching for faster, flexible and easy application development & deployment cycles. Because the PaaS technology solution allows a rapid application development for custom-business apps to be speedily developed & deployed through 'point' and 'click-type' PaaS tools available in web browsers.

According to survey 85% of respondents stated that their organization demands to reduce cost and time (resource consumption) to develop & deploy applications. In this scenario, 70% of the respondents said that they are either already adopted PaaS technologies or planned to use them in near future. 54% of existing users of PaaS reported that PaaS technology helped them to reduce the application development time. While 51% stated that a great deal of costs reduction. 47% respondents said that PaaS had offered more innovation and offered better and enhanced way for technology development [36-38].

Amazon Web Services are basic market leader in IaaS. However, it also stepped-in the PaaS market and has been enhancing its PaaS services. Amazon Web Services has already started potential enhancements with the most recent launch of its Redshift data warehousing cloud service. The Amazon also introduced a Java VM-hosting PaaS based platform that optimized for the cloud application and services. Heroku and Force.com are also well established early pure PaaS vendors. Another leading market service provider in PaaS is Microsoft who has incorporated Microsoft Azure application programming interfaces (APIs) feature to coming Windows Server. It will facilitate third parties to host Azure compatible clouds application. Red-Hat recently initiated PaaS technology products by the launch of "OpenShift", a PaaS technology infrastructure. These all new advancements in technology facilitate cloud migration as well as encourage developers and businesses to rely and move to PaaS to harvest its great capabilities [39]. Further details about these PaaS service providers available in coming sections.

# 5 PaaS Core Architectural Transformations

## 5.1 PaaS Generations

Since the PaaS technology market breakthrough, it has seen many transformations. Now, the evolution of PaaS cloud is moving towards container based application development or interoperable PaaS. The below part outlines some major transformations in PaaS technology and presented in the different generation pattern [13].

The first generation was based on classical fixed proprietary cloud platforms. Major PaaS providers of such technology at that time were Heroku or Azure.

The second generation was developed around open source solutions. In this generation open source PaaS providers like OpenShift and Cloud Foundry allowed clients to run their own PaaS (in the cloud or on-premise) which is already developed around containers based technology. Now, Openshift transferred from its own container model to the Docker based container model. The same transformation performed by Cloud Foundry through its internal Diego solution.

The present and third generation comprises platforms like Deis, Dawn, Octohost, Flynn and Tsuru which are built on Docker based container models. These platforms build around Docker from scratch and are deployable on public IaaS clouds or on own servers.

## 5.2 Containers Revolution

The cloud computing architecture depends on virtualization techniques to accomplish elasticity and productivity for large-scale resource sharing. Up-to last few years, Virtual Machines (VMs) are the most popular way to achieve virtualization in the cloud. However, the idea of containers supported OS-level virtualization is changing the cloud industry significantly. The below section discussed the two basic architectures for the PaaS cloud. This study has also assessed that which architecture is more promising to transform the PaaS industry in the coming future.

Virtual machines (VMs) are considered as backbone infrastructure for operating system layer virtualization. On the other hand, the new idea of "Container" is same as VMs to offer virtualization (lightweight virtualization); however, with less resources and time consumption. Containers have been proposed as a solution for numerous cloud interoperable application packaging.

Containers and virtual machines are both virtualization tools, though solve different problems. Basically containers are PaaS focused tool for delivering application with more and more interoperability and portability. Containers utilize fundamental operating systems (OS) virtualization principles. In contrast, virtual machines are more concerned about hardware management and allocation with the focus on hardware virtualization [13]. Fig-9 shows an architecture level difference among the VMs and containers.

Currently, Containers are being used as an alternative to operating system level virtualization to execute multiple isolated systems on a single host. Containers technology within the single OS is much more effective and with this effectiveness, they have strengthened the future of the cloud infrastructure, more specifically, in PaaS industry through replacing virtual machine architecture [14].

A container is a light weight OS runtime environment inside the host system. It executes instructions native to the core CPU, eradicating the requirements for instruction level emulation or just in time compilation. The key advantage of containers over virtual machines is that they offer resource savings without overhead of virtualization as well as offers isolation [15]. Containers have long existence as extensions to Linux distributions while there are no native commercial containers for Windows yet.

Docker open source containers emerged over the past few years and developed as de-facto standard intended for applications extension form one platform to another. It is also executing as micro-services in Open-Shift and Cloud Foundry PaaS environment. Docker containers have currently became available with key Linux distribution as well as supported in major cloud services.

Currently every major cloud vendor and enterprise infrastructure company has jumped to the Docker trend including IBM Corp., Google Inc., Microsoft, RedHat Inc., VMware Inc., and Rackspace. RedHat is presently, the top outside-open-source contributor for container development. All the latest distribution of RedHat Enterprise Linux offers the Docker containers [16].

Docker containers are playing an important role in transforming the PaaS industry. Now, Docker and related lightweight containers intended to revolutionize the role of operating system and the VMs. For instance, the VM has revolutionized the physical bare-metal server infrastructure. Docker containers are getting popular in PaaS industry because they offer less overhead and better interoperability as compared to traditional VMs.

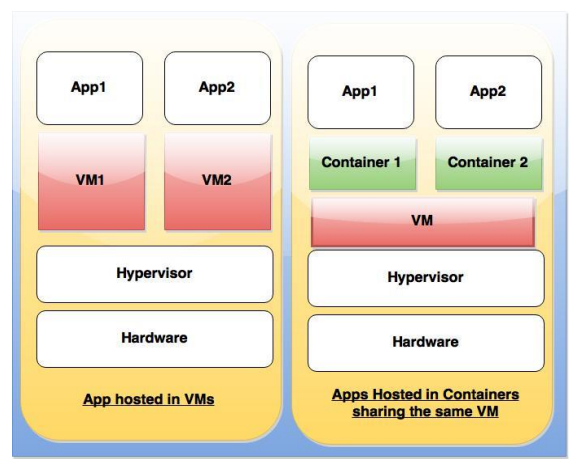

**Fig. 9** VMs v/s Containers

## 6 PaaS Platform Rivalry

Adopting PaaS technology is a very big challenge. When analyzing and selecting a PaaS technology provider, every business or individual have to consider the server side technologies, programming languages and data storage options. Technical support for application integration and developer tools is also a significant factor. Finally, the most important factor that needs to be considered is 'cost'. It is important to assess the cost of running applications in a PaaS platform and need to know the pricing model of a chosen PaaS vendor [17].

This section of paper will discuss the PaaS Platforms from different point of views. Initially, it will present top vendors of PaaS technology with their technologies, potential features, possible issues and business strategies through SWOT (strengths, weaknesses, opportunities and threats) analysis. These top vendors will also be discussed regarding their future market scope. How they plan to capture a maximum market in the near future? The next part will outline some of the up-coming PaaS technology providers with great services, big investments and innovative tools. The basic objective of this study is to discuss these upcoming technology providers and to assess the possible future market stake among already established business giants. The third part of the discussion is "Open Source PaaS". This section will outline possible well performing open source PaaS technology provider and their future. Another latest PaaS paradigm is "Mobile PaaS (mPaaS)" which is in its early stage of development. The mPaaS section will highlight this technology paradigm and associated businesses.

### 6.1 PaaS Market Giants

#### 6.1.1 Amazon AWS PaaS

Amazon is presently considered as the biggest cloud service provider. Amazon Web Services emerged primarily as an IaaS, however, in last few years company also stepped-in in the PaaS through its AWS platform. AWS allows to utilize platform services and capabilities without maintaining or creating own application servers. AWS offers support to a large number of languages like Python, Java, Perl, Ruby and other languages. MySQL, Oracle and SQL Server can be deployed and managed. The Amazon' RDS web services allows to eliminate database administration issues. AWS offer other services like Amazon Elastic Beanstalk for auto-scaling, application health monitoring and automatic load balancing, which are so much supportive and helpful for application developers.

Strengths: As AWS is mainly an IaaS technology provider, so virtually, there is no limit to the databases, languages or server side technologies. We can have capabilities to install and run any of them [18].

Weaknesses: As compared to other PaaS options, AWS platform requires more management overhead which leads to more time consumption tasks.

Opportunities: AWS's Free Usage Tier allows new users to have free of cost 30GB data space, 750 hours and 15GB bandwidth [19].

Threats: Cloud "Locked-in" situation. Also it is very expensive for basic offers.

Pricing Strategy: A Monthly calculator is available to calculate the charges for application services, storage and data charges. However, the overall structure is pretty complex [20]

Future Prospects: Amazon AWS PaaS is progressing continuously to huge scale for getting a better competitive edge in the future. In this scenario, James Hamilton, Vice President and distinguished Engineer at the Amazon Corp., outlined some of outstanding AWS innovation statistics. According to him, there are one million active customers, presently top 14 cloud service providers collectively have 1/5th capacity of AWS (Gartner statistics -2013), 449 latest services and key features released in 2014, continually adding new server capacity for Amazon's global infrastructure to achieve the $7B annual revenue, estimated 132% data transfer growth and have achieved 02Tbps data-center network capacity [21].

AWS is trying to enhance its 'Availability Zone' capability in the coming future for better PaaS services provision. AWS is also trying to establish the Availability Zones all over the world with more and more proximity to clients to minimize messaging latency. This will enhance the state updates synchronously between availability zones, which significantly improves availability in contrast to typical way, where data-centers are pretty much far apart [21].

According to Bermudez, Traverso, Mellia, & Munafo [49] in coming future, AWS PaaS is also trying to enhance services and attempting to avoid public cloud lock-in situations. In an interview with Gigaom Research 2014 in San Francisco: Werner Vogels, Amazon Web Services CTO stated that "we are well aware regarding companies concerns for adoption of public cloud services while we are trying to get customers into the cloud through mutual cooperation with major competitors [22]. It leads to signs that AWS is trying to avoid cloud lock-in situation. In the coming future, it is expected that businesses may not hesitate to adopt public cloud and might not face cloud lock-in situations.

#### 6.1.2 Google App Engine

Google App Engine is one of the major market stakeholders in PaaS services. It is designed for developers who program in languages like Python, Java, Go and PHP for developing distributed web applications. Google App Engine has Java based environment that supports other languages using Java Runtime Environment (JRE). It also has a software development kit (SDK) for each of four main supported languages and Eclipse plug-in. According to Ciurana [51], GAE PaaS supports runtime environment and managing infrastructure that is ensured to scale up and down, however, if the applications conform to the standards of Google App Engine. It also offer schema-less data-store that deals with the complex management of data which is available for several machine instances.

Strengths: Sandbox model allows Google App Engine for isolating the processes from each other. In this way, a risk of process disruption has been reduced (e.g. a rogue process on a physical server, disrupt other process's operations on that server).

Weaknesses: Google App Engine only offers a limited number of programming languages support (Java, Python, Go and PHP).

Opportunities: For new users it provides free first 28 instance hours, cloud storage of 5GB is free, 1 GB outgoing and incoming traffic is also free [23].

Threats: GAE has problem of cloud locked -in.

Pricing: Google App Engine is using the pricing model of pay per use. It charges $0.05/ hour fee for reserved instance or $0.08/hour on demand instance. Cloud Storage service is also available at $0.026/GB/month. The Data-store cost $0.06/100k read/ write operations. While small operations are free. Outgoing Network Traffic (Bandwidth) costs $0.12/GB [23].

Future Prospects: In this IT based age information is power. In case of "Big data" Google has no competitors. In machine learning and data mining field, Google is recognized as one of the top companies in the world. There are huge chances in future that Google Cloud Platform and its services will mostly be dedicated for the Big Data industry (BigQuery).

Google Compute Engine could be the biggest competitor of AWS during 2015 and in coming years. Google has successfully managed and enhanced its cloud infrastructure since last 10 years. Since 2008, Google App Engine has improved up to a great extent. Google is taking full advantage of its global and comprehensive network as compared to AWS [24]. According to Severance [54], Google has worldwide comprehensive network of data-centers. So, Google enjoys a better competitive edge of 'Availability Zone' capabilities.

Google has the future competitive advantage. Unlike the Amazon, Google has more huge data center infrastructure over the globe and has faster market access when it comes to launch PaaS services to new regions around the world. Such aspects will have an immense impact in terms of price and market shares in upcoming future [25].

Google App Engine recently introduced a new feature for its PaaS platform to bridge the gap of communication in global development infrastructure. Google released a beta version of Google Cloud Pub/Sub; a way to link and communicate among services and applications. This will offer Google a huge competitive advantage in the coming future over the other PaaS providers to link services and applications whether they are hosted through Google Cloud Platform or on-premises infrastructure. This app will offer capability to all customers to send up to 10,000 topics and 10,000 messages/ second [26].

### 6.1.3 Microsoft Azure

Microsoft Widows Azure was first released on Feb 1, 2010. In 2014, it was named Microsoft Azure. According to Gartner, Microsoft Azure is one of the major cloud platforms and ranked as a leader in both IaaS and PaaS cloud industry [27]. Now lines blurring between PaaS and IaaS technology of Microsoft. Azure Cloud facilitates languages such as Node.js, .NET, Python, PHP, Ruby and Java. Developers are able to make use of Visual Studio for developing and deploying applications. Also developers have the choice of selecting among SQL Database, Tables and Blobs when persistent storage is required. Microsoft Azure' born applications is administered by its dashboard or by using a command line interface [28].

Strengths: Microsoft Azure is offering PaaS and IaaS in one package. Azure facilitates a quick scale up or down according to the changing needs of business [29].

Weaknesses: One of the key aspects that seem to be bottleneck for Azure is its minimalist administration control portal.

Opportunities: MS Azure offers a free 30-day trial version for new users. This version has a limit of up to $200 [30, 31].

Threats: Higher self-hosting and integration costs.

Pricing: Azure pricing is based on the size of instances running. Price ranges from $0.02/hr. to $0.64/hr. Also outbound data transfers start from the $0.087 per GB. SQL database services start from the $4.99/month. Virtual Machine starts from $13/month [30].

Future Prospects: Microsoft Azure turned out to be a major market competitor with Google and AWS in recent years. To get its distinguishing market position, Microsoft Azure introduced a number of innovative and enhanced services, however, the future lies to technology providers who offer innovation. In this scenario, Microsoft Azure is trying to break-down the boundaries among PaaS cloud services and that's why company delivers a consistent hybrid PaaS cloud services. In future, Microsoft Azure is aimed to offer services for enhancing application mobility, easier development and avoiding public cloud lock-ins [32].

At the GigaOM conference in San Francisco June 18–19, 2014, Scott Guthrie, the Executive Vice President of Microsoft's Cloud and Enterprise group, stated that "Microsoft achieved hyper-scale cloud service providers' status along with Google and Amazon. In coming future, Microsoft will differentiate itself to customers through offering a huge variety of services (from high-level products to raw access to server hardware)" [33].

In case of getting better future market competitive edge, in July 2014, Microsoft Azure released Azure Machine Learning (ML) capability. According to Joseph Sirosh, Microsoft Corporate Vice President of Machine Learning, stated that a comprehensive set of services and capabilities of machine learning added at Microsoft Azure. In future, these machine learning tools will offer the ability to forecast business trends by reducing forecast times for customers and partners from months to hours [34, 35].

Sirosh added that customary approaches for data analysis help to predict the future. But the new machine learning capability will change the future. Microsoft Azure expected to attain a huge market in the coming future through innovation of services, as Azure ML will help business to detect patterns, predict disease outbreaks, forecast demand and prevent crime etc. [35].

6.1.4 Salesforce.com & Heroku

Salesforce.com is a key PaaS service provider. Currently it is offering PaaS services through Force.com and Salesforce1.com. Force.com has established standard for developing new age multitenant cloud applications. Force.com offered PaaS infrastructure to developers to develop and deploy the application on Salesforce's servers. Salesforce.com PaaS platform offers support for programming languages and frameworks like Java, Apex, Ruby on Rails, Node.js, Python, all JVM languages and more. Such extensive support offer developers to develop all kinds of rich and effective applications that please customers [36].

Heroku: Heroku is one of the initial cloud platforms, founded in 2007. Salesforce.com has acquired company in 2010; though it is still working as a subsidiary cloud service. It supports languages and frameworks like Python, Ruby, Scala, Java, Node.js and Cloture. It is based on the abstract computing infrastructure known as dynos, which run processes in an isolated environment, based on virtualized Unix-style containers [37]. Heroku perform very efficiently with apps that support Twelve-Factor-App methodology. Heroku is now becoming more and more mature PaaS platform with numerous better capabilities (support third party apps e.g. Add-ons). In future, it will become a major market shareholder of PaaS [37, 38].

Here Heroku and Salesforce will be considered as one entity regarding SWOT analysis and future prospects.

Strengths: Salesforce.com PaaS offers a pre-integrated capability that already has a search database, business intelligence reports and security/identity etc. Business services also manage upgrades and automatically back-up of user data [17, 39].

Weaknesses: Salesforce.com PaaS services charge very high recurring subscription costs as compared to other market competitors.

Opportunities: The initial pricing package of Salesforce.com PaaS offers powerful applications with access to ten custom objects/users [40].

Threats: Data center's reliability is questionable due to a number of significant outages.

Pricing: Salesforce.com PaaS pricing is based on monthly basis. The starting cost is $25/user/month. Pricing is billed on annual basis [40].

Future Prospects: Salesforce1 initiated PaaS services through Force.com platform. For gaining a better competitive edge in the market, the company took a lot of initiatives. For example, till the July 2014 company acquired 31 different companies (acquired Heroku PaaS Platform; in Dec 2010, acquired ExactTarget a Marketing Cloud company in June 2013, now known as Salesforce Marketing Cloud, etc.). This aspect shows that Salesforce.com is really determined for its future and getting top position in cloud and PaaS technology. Force.com & Heroku collectively supporting Salesforce.com mission for a bright future and bring together the power of both into the one family of PaaS. These both companies offer complete and comprehensive set of PaaS platform and tools [41-43].

A novel technology framework is going to release in spring 2015 for developing better user interfaces and lightning components. Using the open source Aura Framework, Lightning components will support for Apex as a server side language and as a replacement for Aura's Java dependency [44].

The Lightning technology framework will be a collection of tools and technologies that will significantly upgrade Salesforce1 Platform. Lightning platform includes components and extensions that permit to develop reusable components, standalone apps and customize the Salesforce1 Mobile App. It will also be used as a new user interface tool that permits to develop apps lightning fast through components offered by platform developers and Salesforce. In future, this tool will support for visualizing, building and automating business processes. It will also work as Schema Builder for viewing fields, developing objects and relationships. This tool is expected to enhance the developer's capabilities to a greater extent and reduces the complexity factor [45].

6.2 PaaS Providers: on the Horizon

6.2.1 IBM's BlueMix

Bluemix is based on Cloud Foundry architecture that enables us to rapidly develop, deploy and manage our cloud applications. Because IBM Bluemix is based on Cloud Foundry open source architecture, so, we are able to tap into a growing ecosystem of runtime services and frameworks. IBM's BlueMix PaaS platform was launched as an open beta back in Feb 2014 and it has proved very successful. BlueMix is presently accessible through IBM's cloud hosting platform 'SoftLayer'. BlueMix presents developers a simple way to develop cloud applications for businesses. BlueMix PaaS supports a number of languages like Node.js, Java, PHP, Go, Ruby and Scala. Also support frameworks like Sinatra and Ruby on Rails. IBM's BlueMix is a fast growing PaaS platform and having a great deal of future for cloud computing and also for business competitive edge [46-48].

Future Prospects: "IBM is leading to a new age of innovation through partnering with technology developers in an open environment to speed-up the evolving world of hybrid cloud computing," said by Robert LeBlanc, IBM's senior vice president of software and cloud Solutions. According to Pattathe, Hoang, Yuen, & Li [79] IBM's BlueMix PaaS platform is progressing speedily into a new era of cloud developing. In a very small time, company attained a very considerable place in the cloud market. For becoming more vibrant for future PaaS industry, IBM is taking a number of steps involving the acquisition of new companies and alliance with world big technology firms. In March 2014, BlueMix and IBM acquired DBaaS firm Cloudant. Now Cloudant turned out to be a core component of the BlueMix PaaS services [49, 50].

Another big step toward IBM's BlueMix future is the recent Apple and IBM partnership. "This partnership will offer capability to IBM for selling industry specific applications which are pre-loaded on iOS supported devices. This will probably fuel further demand and growth of mobile apps. "This technology exchange alliance with Apple business will offer a boost in bringing innovations to cloud clients worldwide. It will also enhance IBM's management for software services and cloud" said by Ginni Rometty. (IBM's chairman, Cupertino, California, 15 Jul 2014) [49, 51].

IBM's BlueMix recently incorporated Watson Analytics (a natural-language based cognitive service) to its stack of cloud services. This tool will help the PaaS apps developers to build powerful analytical solutions. These solutions will help businesses to analyze, aggregate and visualize huge amount of data, and information to uncover quality and unique insights.

There are lots of tools including Watson, DashDB, Cloudant, DataWorks, and other high quality tools are also available at IBM' Bluemix stack. It is expected that developers will get an enormous power and capability to build next generation PaaS born apps [52].

6.2.2 AppFog

AppFog is another leading PaaS provider. It offers support for a multi-language and multi-framework PaaS. It offers a great capability for multi-cloud deployments together with private clouds. AppFog offers support for languages and platforms like Ruby, Java, Node, Python, PHP, Scala and Erlang. As well as it offers PostgreSQL, MySQL, RabbitMQ and Redis along with 3rd party add-on services. AppFog is designed around the open source Cloud Foundry architecture and facilitates SVN, Mercurial and Git for code management [53].

The biggest advantage of AppFog is that it can run applications on multiple clouds. People who prefer the VMware based architecture will definitely prefer PaaS. For initial users, AppFog offers free version with unlimited apps within 100MB persistent storage (supporting PostgreSQL and MySQL), 2GB of RAM and up-to 8 service instances. The Overall pricing system is based on pay for only what it is used. A big issue with AppFog platform is that it does not have a persistent file system. So if another object storage system is needed; it will be integrated but at an additional cost [53, 54].

6.3 PaaS Open Source Padrone

6.3.1 Cloud Foundry

Cloud Foundry emerged in year 2008 as an open source PaaS cloud computing platform. It is a Java based PaaS platform. It was originally developed by VMware. Later it was acquired by Pivotal Software. Cloud Foundry support languages like Scala, Ruby, Python, PHP, Node.js, Java and platforms like Play 2.x, Lift, Rails, Sinatra, Spring Framework 3.x, 4.x. In February 2014, VMware declared the establishment of the Cloud Foundry Foundation, with Platinum sponsorship of EMC, Pivotal, Rackspace, IBM and VMware as Platinum sponsors. Cloud Foundry Foundation is having 33 members and 42 contributing development businesses.

Cloud Foundry obtains source code from Ruby users and developers. It is an open source PaaS platform that permits deployment of apps to Amazon Web Services (AWS), OpenStack, vCloud Air, vSphere and vCloud Director. The key hosted services offered by Cloud Foundry are MongoDB, MySQL and RabbitMQ. Cloud Foundry' PaaS developers get a great deal of support and ease by having tools like the Eclipse Plugin, command line tools, application scale tools and a build integration tools. These days Cloud Foundry is heavily competing against the Heroku, AppScale and OpenShift [55]. Recently on Dec 9, 2014 Linux Foundation Collaborative Project took Cloud Foundry as a collaborative project. So, in future, company expected to be a major competitor for the other open source PaaS companies [56, 57].

### 6.3.2 Engine-Yard

Engine-Yard is one of the early cloud service providers. Now it is getting pace. One of the main reasons is its strategic alliance with Microsoft in 2013. Through this alliance, developers will be able to use its open source PaaS capabilities running on Microsoft cloud infrastructure. Engine-Yard is offering support to languages like PHP, Ruby on Rails and Node.js. Engine-Yard is offering a great deal of capability for operations management, snapshot management, backup, cluster handling, load balancing and administering the database [58].

Engine-Yard is trying to be more and more vibrant in the open source cloud market. For this, the company has offered lots of innovative services and launched a number of collaborative projects with the world's well-known technology companies. As mentioned above, the strategic alliance of Microsoft and Engine Yard will deliver commercial grade applications in the future [59].

Another major development for Engine-Yard is its alliance with Oracle Crop. Oracle took Engine Yard technology services for development of Oracle Cloud PaaS. Engine Yard PaaS automates the deployment, configuration and on-going maintenance of Oracle PaaS to facilitate developers [60, 61].

### 6.3.3 Red-Hat's OpenShift

Red-Hat's OpenShift is PaaS based platform that is based on open source applications. OpenShift presents a huge variety support to databases, languages and components. OpenShift PaaS technology platform is vastly customizable and available in three forms. At first, OpenShift Enterprise is a private PaaS that support and manage client data center. Secondly OpenShift Online is a PaaS cloud based hosting service. Finally, OpenShift Origin is an open source application hosting site [62]. OpenShift offers a lot of new features for better cloud development like it automates system management jobs. For example, configuration and scaling, virtual server provisioning and supports GIT repositories intended for code management [63, 64].

In November 2014, Red Hat, Inc. released OpenShift Enterprise 2.2; a new technology based platform for strengthened its early PaaS OpenShift services. This new technology service release offers a new private integration PaaS Service (iPaaS). Such technology development based initiatives shows that Red Hat, Inc. is determined to obtain an outstanding position in the open source cloud market. Integration of iPaaS will offer services to future developers for application messaging and integration. Such messaging and integration tools (JBoss A-MQ for xPaaS & JBoss Fuse for xPaaS) will support for easy and speedy integration that will lead to increase application time-to-market [65].

### 6.3.4 Stackato

Stackato is a stable, secure and commercially supported PaaS platform which is built through various top open source packages like Docker and Cloud Foundry [66]. ActiveState's Stackato is a new technology based platform that offers support to develop and deploy new era web apps. Stackato offers flexibility and ease, supported by direct virtual machine (VM) access on IaaS. It also offers highly automated configuration that support developers to connect PaaS with IaaS [67]. Stackato services include a web management console, a customizable app store, activity stream as well as self-service. There are also some other features for PaaS developers including auto-configuration, end-to-end development, dynamic load balancing, elastic scalability, a centralized cluster administration, application auto-scaling, placement and availability zones, as well as persistent file-system sharing [68, 69].

Stackato released new version Stackato 3.0 on November 2013. This new version holds the support of Docker. After Docker alliance with Stackato application development, business gained incredible momentum. Docker permits efficient and powerful portability and container management across Stackato clusters. With minor updates and bug fixations, a latest 3.0.8 version (released in May 2014) is available now. Newer versions expected to bring more power and support for developers and will offer Stackato a distinguishing place in PaaS market [70, 71].

### 6.3.5 Cloudify

Cloudify (by GigaSpaces) is an emerging PaaS technology platform that acts as an orchestrator for applications on the cloud. Developers can deploy, manage and scale the complete application stack through tools offered by Cloudify [72]. The Cloudify manager deals the processes through monitoring different states of the application as well as deploys the program agents. Cloudify's open architecture permits clients to develop and utilize custom workflows, data types and plugins to develop a model that works only for developer [73, 74].

Cloudify is constantly working and improving its services and tools for better future. The Cloudify 3.1 is the latest version which was published on Dec 17, 2014. It offers novel features and tools for PaaS developers for better, easy and fast development [75].

### 6.4 Mobile PaaS

In last few years, some great PaaS services have emerged to support and facilitate the developers. These PaaS cloud services have taken away many time-consuming and hard phases of application development. However, for mobile developers, there is still something to be desired. Here comes the Mobile Platform-as-a-Service (mPaaS). mPaaS offers powerful tools and services to rapidly develop, integrate, deploy and manage data driven mobile applications for mobilizing workflows, processes, reports, tasks, databases and custom applications [76, 77].

mPaaS is an emerging arena of PaaS technology. Many new players are stepping-in to get maximum market share. This section is about the top mPaaS provider in the market.

### 6.4.1 mFicient

The mficient is an mPaaS platform aimed to speed up and improve the deployment of business workflows and processes on mobile devices. This platform offers device and app management, rapid development tools, single sign-on integration with Active Directory (LDAP), easy backend data integration and without complications of software or hardware installation etc. [78].

### 6.4.2 Kumulos

Kumulos is an original Mobile Backend as a service (MBaaS) platform developed for mobile application developers. Mobile Backend as a service (MBaaS) is a cloud model which offers capability to mobile application developers with a method to link their apps to backend APIs and cloud storage. Kumulos allow mobile app developers to focus on developing a high class and appealing app experience. This technology platform offers easy integration of sophisticated features and existing data sources into app. Kumulos work on iOS & OSX, Windows Phone, Android, Blackberry and HTML 5 [79].

### 6.4.3 Kinvey

Another market leader in mPaaS is Kinvey, which offers a platform to develop digital business apps with faster mobile Backend as a Service. It also offers capability to deliver apps faster by means of utilizing development resources of client choice. Kinvey offer support capability to develop any kind of application for any kind of mobile device with no platform lock-in problem. It offers safe mobile data and identity, standardize backend integration as well as scale up or down through turn-on infrastructure [80].

## 7 PaaA' Prospects

Year 2015, named as "the year of PaaS". Even PaaS technology was around since the evolution of cloud computing. Also the PaaS field saw its initial industry validation in 2010, when Salesforce acquired Heroku. Huge transformations happened since that time. However, PaaS came to the front now because the concept of PaaS or its awareness, people recognize it today; they were not recognized 4-5 years ago. Another reason can be that since the establishment of cloud technology, people were so much concerned about security issues. However, most of the issues are being resolved and now people starting believing cloud as a reliable technology platform. Another aspect is that PaaS was bottlenecked behind the debate of IaaS cloud computing because platform applications require an IaaS infrastructure. But today most of IaaS issues and infrastructures aspects are being settled. Therefore, PaaS is in the position to move on.

Another aspect of current PaaS technology hype is that presently, the demand for applications is skyrocketing. This presents a great pressure on IT firms to develop and deploy applications quickly. In such situation, IT firms are seeking new platforms, tools and procedures to support them to move from slow-moving manual application development practices to automated procedures. In this situation, PaaS is the best candidate for fast, low cost and easy development and deployment.

Another factor that accelerate the PaaS acceptance is the open source licensing. PaaS vendors like Cloud Foundry, revolutionize the PaaS application distribution, which offers an open source-based product that can be deployed virtually at any infrastructure. This makes PaaS a much accepted and preferred choice for IT enterprise. Now IT firms are feeling free to pursue the high productivity and flexibility that PaaS offers.

PaaS is being accepted as a reliable development platform. It is also predicted that PaaS will achieve huge milestones in near future. Though, businesses still confused between private, public or hybrid PaaS. As already stated, that choice between public, private and hybrid involves a lot of factors including cost, privacy, reliability, freedom, technological support and packages etc. In this scenario, Dyn global leader firm in internet performance performed the fourth annual survey on the "Future of Cloud Computing" (Manchester, NH, June 2014). This research is performed in alliance with Gigaom Research, North Bridge Venture

Partners and other 70 collaborating organizations. The result of this survey offers a glimpse into next generation cloud computing [81]. This survey showed that 42% of companies presently emphasizing on adoption of hybrid clouds, while it is expected that it will increase to 55% in the next 5 years [32]. As the Hybrid cloud is turning out to be an ultimate choice of businesses. Till now, it is seen that in the hybrid cloud sector, Amazon and Salesforce.com are performing very well. In the future, it is expected that these companies will capture a maximum market share [82]. Hybrid cloud offers more reliability, the best features of public and private PaaS so Hybrid cloud can be the future of PaaS cloud development.

Choosing the right PaaS platform for business is really important for any organization. It involves analysis of number of factors and technological aspects. One of the main aspects is cloud basic architecture. Traditional PaaS platforms are based on VMs supported architecture. However, containers changed the whole game. Docker and Linux containers are poised to totally transform the methodology of application development, deployment and instantiated. The idea of containers accelerated the way of application delivery through making it easy to package app along with their dependencies. Consequently, the similar container based app is able to execute in different development, test and production environments.

We can say that era of containers based application development has arrived. Application packaging is becoming more significant because of the arrival of fog computing, Internet of Everything and a return to decentralized systems.

Cloud vendors are well responding with containers based innovations. Many cloud vendors initiated high-profile projects including Apache Mesos, Red Hat OpenShift, Google Kubernetes, VMware CloudFoundry and Cisco application containers [83].

Beside the new evolving PaaS paradigms, fundamental PaaS architecture is also transforming. Mobile PaaS is also an emerging PaaS paradigm that promises a great future. It is in evolutionary stages and need more time to mature. mPaaS also has great potential to grab the huge market share because in future, the demand for mobile apps is going to be higher and higher.

Future work is about analysis of Hybrid PaaS and its possible assessment of the success factor for a business. This work will also involve statistical assessment of performance and gains attained through this new technology platform.

## 8 Conclusion

Imagine it, PaaS will build it. This term is absolutely true about present PaaS technology. The next generation of PaaS will be achieving the real promise of object-oriented development and 4GLs. Rapid development with less cost and work is now becoming reality. PaaS transformed traditional application development approaches. Now development is bit faster, flexible and cheaper. This happens due to the elimination in most of infrastructure related tasks.

In future, PaaS clients will experience much deeper levels of abstraction. Application development within PaaS technology will be faster, less costly and will offer better quality. The Next-generation PaaS cloud is expected to revolutionize the development field through offering ease to non-programmers and non-techies to develop the latest software applications at high speed for real competitive benefits. Businesses will also be able to focus on outlining their software requirements intuitively and logically for more concrete developments. Business stakeholders will make use of PaaS cloud to standardize their own quality software.

This research has assessed and analyzed the major trends, market status and future aspects of PaaS cloud. It also outlines the present market situation and future trends of PaaS technology. Top PaaS market stakeholders and technology providers have been discussed with possible potentials to assess their possible future in the emerging PaaS market. Salesforce.com and Amazon AWS PaaS is performing quite well in the current market. Heroku, Google App Engine and Microsoft Azure are trying to become more and more competitive in the next few years. While from upcoming PaaS companies, IBM's BlueMix is one of the top vendors, who initiated business with huge investment and excellent quality services. Open source PaaS platforms are also performing quite well and now offers a great deal of competition to proprietary cloud providers. Cloud Foundry and Red-Hat's OpenShift is one of the key examples of it.

In the coming future, PaaS technology provider will grab the major market share, which will offer more language support, automated development management tools, non-platform-lock-in development environment, security, quality of services and the most important low cost services. Container technology is one of the key examples of PaaS innovation. New generation of PaaS application development is container based technology with more ease, abstraction and low resource consumption. Though, few security issues are still there but the future of PaaS is much aligned with container supported technology.

## Acknowledgement

This research is supported by CAS-TWAS and University of Science and Technology, China. We thank all colleagues who provided insight and expertise that greatly assisted our research.